\documentclass[journal=jacsat,manuscript=article]{achemso}

\usepackage[version=3]{mhchem} % Formula subscripts using \ce{}
\usepackage{algorithm}
\usepackage{amsmath}
\usepackage{algpseudocode}
\usepackage{threeparttable}
\author{Jie Li}
\affiliation{ Pitzer Center for Theoretical Chemistry, Department of Chemistry, University of California, Berkeley CA 94720, USA}
\altaffiliation{Equal contribution}

\author{Jiashu Liang}
\affiliation{ Pitzer Center for Theoretical Chemistry, Department of Chemistry, University of California, Berkeley CA 94720, USA}
\altaffiliation{Equal contribution}

\author{Zhe Wang}
\affiliation{ Pitzer Center for Theoretical Chemistry, Department of Chemistry, University of California, Berkeley CA 94720, USA}

\author{Aleksandra L. Ptaszek}
\affiliation{ Pitzer Center for Theoretical Chemistry, Department of Chemistry, University of California, Berkeley CA 94720, USA}
\affiliation{Christian Doppler Laboratory for High-Content Structural Biology and Biotechnology, Department of Structural and Computational Biology, Max Perutz Labs, University of Vienna, Campus Vienna Biocenter 5, 1030-Vienna, Austria}
\affiliation{Laboratory for Computer-Aided Molecular Design, Division of Medicinal Chemistry, Otto Loewi Research Center, Medical University Graz, Neue Stiftingtalstrasse 6/III, 8010-Graz, Austria}

\author{Xiao Liu}
\affiliation{ Pitzer Center for Theoretical Chemistry, Department of Chemistry, University of California, Berkeley CA 94720, USA}

\author{Brad Ganoe}
\affiliation{ Pitzer Center for Theoretical Chemistry, Department of Chemistry, University of California, Berkeley CA 94720, USA}

\author{Martin Head-Gordon}
\affiliation{ Pitzer Center for Theoretical Chemistry, Department of Chemistry, University of California, Berkeley CA 94720, USA}
\affiliation{Chemical Sciences Division, Lawrence Berkeley
National Laboratory, Berkeley, California 94720, United
States;}

\author{Teresa Head-Gordon}
\affiliation{ Pitzer Center for Theoretical Chemistry, Department of Chemistry, University of California, Berkeley CA 94720, USA}
\affiliation{Departments of Bioengineering and Chemical and Biomolecular Engineering, University of California, Berkeley, CA, USA}
\email{thg@berkeley.edu}

\title
  {Highly Accurate Prediction of NMR Chemical Shifts from Low-Level Quantum Mechanics Calculations Using Machine Learning}

\abbreviations{IR,NMR,UV}

\begin{document}

\begin{tocentry}

\begin{figure}[H]
\begin{center}
     \includegraphics[width=0.98\textwidth]{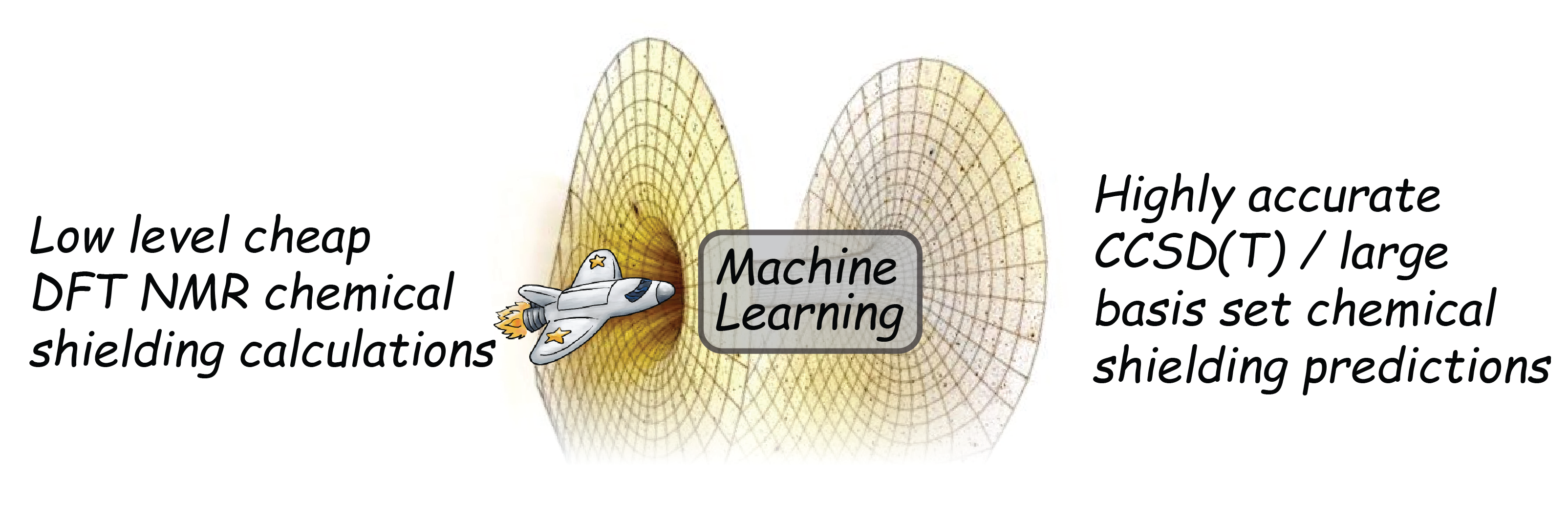}
\end{center}
   
    \label{fig:TOC}
\end{figure}

\end{tocentry}

\newpage
\begin{abstract}
\noindent
Theoretical predictions of NMR chemical shifts from first-principles can greatly facilitate experimental interpretation and structure identification. However, accurate prediction of chemical shifts using the best coupled cluster methods can be prohibitively expensive for systems larger than ten to twenty non-hydrogen atoms on today's computers. By contrast machine learning methods offer inexpensive alternatives but are hampered by generalization to molecules outside the original training set. Here we propose a novel machine learning feature representation informed by intermediate calculations of atomic chemical shielding tensors within a molecular environment using an inexpensive quantum mechanics method, and training it to predict NMR chemical shieldings of a high-level composite theory that is comparable to CCSD(T) in the complete basis set limit. The inexpensive shift machine learning (iShiftML) algorithm is trained through a new progressive active learning workflow that reduces the total number of expensive calculations required when constructing the dataset, while allowing the model to continuously improve on data it has never seen. Furthermore, we show that the error estimations from our model correlate quite well with actual errors to provide confidence values on new predictions. We illustrate the predictive capacity of iShiftML across gas phase experimental chemical shifts for small organic molecules and much larger and more complex natural products in which we can accurately differentiate between subtle diastereomers based on chemical shift assignments.  
\end{abstract}

\section{Introduction}
Nuclear magnetic resonance (NMR) spectroscopy is a highly accurate experimental technique to probe chemical bonding and subtle environmental differences of atoms in various molecular systems, ranging from small molecules\cite{jacobsen2016nmr, NMR_book,modern_NMR}, natural products\cite{jacobsen2016nmr,bagno2006toward,saielli2011addressing}, biopolymers\cite{protein_solution_NMR_determination,intro_bio_NMR}, to materials.\cite{brown2012applications,mackenzie2002multinuclear,KRR} The NMR chemical shift (CS), which describes the shielding effect offset of a nucleus of interest relative to a defined standard molecule, is one of the most informative data obtained from an NMR measurement, especially for molecular structure\cite{structure_validation}, identifying the crystal morphology from a selection of candidates\cite{KRR}, distinguishing among synthetic outcomes for natural products,\cite{saielli2011addressing} and building and refining atomic level models for proteins.\cite{protein_refinement}. 
Therefore, accurate CS back-calculators which connect structure to shift perturbations are an indispensable tool in trying to help scientists understand and make good use of NMR chemical shifts measurements.

Chemical shifts arise from the electron shielding of a nucleus under an external magnetic field. The shift values can be calculated from first principles\cite{helgaker1999ab,gauss2002electron,lodewyk2012computational} using the second order magnetic shielding tensor $\hat{\sigma}$, that describes the response of the induced magnetic field in all directions, but usually only the isotropic component $\sigma_{iso}=\frac{1}{3} \mathrm{Tr}(\hat{\sigma})$ is mapped to an experimental observable.\cite{webb2007modern}  Calculation of chemical shifts can be done with exceptional accuracy using  coupled-cluster theory with single and double excitation and perturbative-approximated triple excitations [CCSD(T)] together with a complete basis set (CBS) or one that is sufficiently large for convergence.\cite{CCSDT_NMR,DFT_benchmark_with_CCSD,NS372} However, with present-day algorithms and computing resources, such calculations are essentially impractical for any complex systems that contain more than ten heavy atoms (non-hydrogen atoms), due to their computational scaling. Efforts continue to reduce the cost by approaches such as composite methods\cite{reid2015approximating,levels}, and nucleus-optimized electronic structure models.\cite{wong2023silico}

Alternatively, data-driven approaches have also been quite successful in predicting experimental or calculated chemical shifts at greatly reduced cost. For aqueous proteins, chemical shifts can be predicted from carefully curated features extracted from 3-dimensional geometries of the peptides using machine learning (ML) methods including neural networks and random forests, such as implemented in SPARTA+\cite{sparta+}, SHIFTX2\cite{shiftx2} and UCBShift\cite{UCBShift}. For organic small molecules in crystalline form, kernel ridge regression (KRR) \cite{KRR}and 3D convolutional networks (CNN)\cite{MR3DDenseNet} have been employed to predict chemical shieldings calculated using gauge-including projector-augmented waves (GIPAW) density functional theory (DFT) methods from merely the molecular structure inputs. Recent work by Guan et al. has trained a 3D graph neural network to predict H and C chemical shifts for neutral organic molecules found in NMRShiftDB\cite{Kuhn2015} using quantum mechanics (QM) optimized geometries and DFT calculated chemical shifts, and then transfer learning to predict experimental chemical shifts from force-field optimized geometries.\cite{CASCADIA} These ML models that directly predict chemical shifts from input geometries are orders of magnitude faster than QM calculations, and can usually achieve comparable accuracy to the quantum mechanical method they have been trained on. However, this has typically relied upon DFT that can calculate chemical shieldings at a much more acceptable cost, but also can often suffer from insufficient accuracy\cite{ab_inito_NMR,comparison_model,benchmark}. In addition, machine learning methods are not expected to generalize to a different molecular system, unlike QM methods that are still much more generalizable and rigorous in terms of predicting chemical shieldings for a specific input geometry.

The question arises whether a machine learning method can be used to ``amend" a low-level QM prediction to high accuracy, hence achieving generalizability and speed at the same time. An intuitive way is to use machine learning to predict the difference between a high-level and low-level calculation, using molecular geometries as input. Such $\Delta$-machine learning idea are exemplified in the work of Unzueta, et al. that predicts a correction to a cheap DFT calculation using small basis set and arrives at the target accuracy of the same DFT method with a large basis set.\cite{delta_ML} Very recently, Büning and Grimme have shown that a similar approach can correct DFT predictions of chemical shieldings to CCSD(T) quality, signifying an important step in predicting CS at the highest level of theory achievable from theoretical calculations.\cite{delta_latest}  

But what is true about many such ML approaches is that they can be poor in predicting out-of-distribution cases, i.e. outside the specifics of the training data.\cite{Haghighatlari2020} Ideally, good feature engineering can  provide an augmented chemical representation beyond just molecular configuration\cite{Haghighatlari2020,UCBShift}, information that is preferably derived from a cheap calculation but which is nonetheless invaluable information for not only obtaining high-level accuracy, but transferability.  This very idea has been proven for predicting correlation energies at MP2 and CCSD level using molecular orbital features at the mean-field Hartree-Fock (HF) level\cite{HF_energy_to_CCSD}. 

In this work we present a novel feature representation obtained from a low-level DFT chemical shielding calculation of the diamagnetic (DIA) and paramagnetic (PARA) shielding tensor elements, and combine it with geometric-dependent features that are used as input into a neural network model to predict chemical shieldings equivalent to CCSD(T)/CBS accuracy.\cite{levels} In addition we introduce a novel active learning (AL) training procedure that selects out-of-distribution training data with increasing number of heavy atoms from a full set of off-equilibrium geometries obtained from the ANI-1 dataset.\cite{ani1} Finally, to analyze the transferability of iShiftML to other systems, we find that error estimations in terms of the standard deviation among a committee of ML models is well correlated with the actual error without knowing the target values, signaling when the model is or is not trustworthy for applications outside the original training set. 

The resulting iShiftML model trained with data up to 7 heavy atoms has exceptional predictive performance when evaluated on the 8 heavy atom test data, achieving prediction errors of 0.11 ppm for H, 1.54 ppm for C, 3.90 ppm for N and 6.33 for O between predicted chemical shieldings and the target CCSD(T) composite method values. The iShiftML model when compared against experimental gas phase CS measurements for molecules that are not included in the training set reduces the error of the low-level DFT calculation by at least 50\%. Furthermore, we have used our method to predict experimental CSs for natural products that are vastly larger and more chemically complex than any molecule from our training dataset, illustrated with strychine and vannusal, in which we show that diastereomers of the vannusal B molecule can be easily differentiated by inspecting the errors between predicted CS and experimental measurements. We expect the iShiftML method to be used extensively to achieve highly accurate predictions of chemical shifts of various molecular systems, and also facilitate theoretical research of NMR chemical shieldings at the CCSD(T)/CBS level.  

\section{Methods and Models}
\subsection{Feature Selection for Machine Learning of Chemical Shifts} 
The magnetic shielding tensor $\hat{\sigma}$ is defined as the total second derivative of the energy $E$ with respect to nuclear spin $\mathbf{M}^A$ at nucleus $A$ and the external magnetic field $\mathbf{B}^{ext}$, with components defined as 
\begin{equation}
%    \hat{\sigma}=\frac{\partial^2E}{\partial \mathbf{M}_{nuc} \partial \mathbf{B}^{ext}}
\sigma_{ a b}
=\left.\frac{\mathrm{d}^2 E(\mathbf{M}^A,\mathbf{B})}{\mathrm{d} M^A_{ a} \mathrm{d} B_b} 
\right|_{\textbf{B}=0, \textbf{M}^A=0} 
\end{equation}
Here ``d'' means total derivative, and $a, b$ correspond to Cartesian indices. For a variationally optimized wavefunction with parameters, $\boldsymbol{\theta}$ (even the exact wavefunction), the total derivative has two partial derivative contributions:
\begin{equation}
\sigma_{ a b} = \left.\left\{ 
\frac{\partial^2 E(\mathbf{M}^A,\mathbf{B})}{\partial M^A_{ a} \partial B_b} 
+ \frac{\partial^2 E(\mathbf{M}^A,\mathbf{B})}{\partial M^A_{ a} \partial \boldsymbol{\theta}} 
\frac{\partial \boldsymbol{\theta}}{\partial B_{ b} }
\right\}\right|_{\textbf{B}=0, \textbf{M}^A=0,\boldsymbol{\theta=\theta}_\mathrm{opt}}
\label{eq:1}
\end{equation}
%where , $P_{\mu \nu}$ represents the one-particle density matrix (with basis functions $\mu, \nu$) and $h_{\mu v}$ is the one-electron hamiltonian matrix. 
Given that the chemical shielding tensor and each of its components, $\sigma_{ a b}$, can also be decomposed into diamagnetic and paramagnetic components within the DFT gauge-including atomic orbitals (GIAO) approach\cite{Wolinski1990,GIAO},
\begin{equation}
\begin{pmatrix}
\sigma_{xx} & \sigma_{xy} & \sigma_{xz}\\
\sigma_{yx} & \sigma_{yy} & \sigma_{yz}\\
\sigma_{zx} & \sigma_{zy} & \sigma_{zz}
\end{pmatrix}
%\label{eq:6}
= \begin{pmatrix}
\text{DIA}_{xx} & \text{DIA}_{xy} & \text{DIA}_{xz}\\
\text{DIA}_{yx} & \text{DIA}_{yy} & \text{DIA}_{yz}\\
\text{DIA}_{zx} & \text{DIA}_{zy} & \text{DIA}_{zz}
\end{pmatrix} + 
\begin{pmatrix}
\text{PARA}_{xx} & \text{PARA}_{xy} & \text{PARA}_{xz}\\
\text{PARA}_{yx} & \text{PARA}_{yy} & \text{PARA}_{yz}\\
\text{PARA}_{zx} & \text{PARA}_{zy} & \text{PARA}_{zz}
\end{pmatrix}
\label{eq:2}
\end{equation}
it comes naturally that the isotropic chemical shieldings at the same level of theory can be calculated as
\begin{equation}
\sigma_{iso}=\frac{1}{3}(\sigma_{xx}+\sigma_{yy}+\sigma{zz})=\frac{1}{3}(\text{DIA}_{xx}+\text{DIA}_{yy}+\text{DIA}_{zz}+\text{PARA}_{xx}+\text{PARA}_{yy}+\text{PARA}_{zz})
\label{eq:3}
\end{equation}
in which the off-diagonal elements have a contribution of zero to the final isotropic chemical shielding formula. However, the full tensor of Eq. \ref{eq:2} still encodes useful information about the local atomic environments for each nucleus and might be helpful with predicting chemical shieldings at a higher level of accuracy. Hence we formulate the chemical shift tensor components DIA and PARA as a feature set for the machine learning approach described further below.

In addition, we use Atomic Environment Vectors (AEVs) as geometric descriptors that are used to describe the atomic environments at each nucleus, following previous studies\cite{ani1, delta_ML}. AEVs are reformulations of the atomic symmetry functions used by Behler and Parinello in their neural networks for predicting molecular energies\cite{behler_parrinello_NN}, which contain orientation-independent angular and radius terms that are determined by local geometries of nearby atoms categorized by atom type within a cutoff. The 384-dimensional AEV for an atom constitutes a radial part (the first 64 elements) and an angular part (the remaining 320 elements). The radial elements for atom $i$ are calculated as
\begin{equation}G_{A,n}^{(rad)}=\sum_{j\in \mathcal{N}[i]}e^{-\eta (R_{ij}-R_n)^2}f_C(R_{ij})
\label{eq:4}
\end{equation}
where $A$ denotes a specific atom type of H, C, N, O for the second atom, and $n$ is a distance index that defines the different reference distances $R_n$ from the center atom. The summation is done over all neighbor atoms $j$ with type $a$ near the central atom $i$ within a cutoff, and $R_{ij}$ is the distance between atoms $i$ and $j$. The reference distances are defined as $R_n=0.9+a_0/2*n$ where $a_0=0.529177$\r{A} is the Bohr radius and $n$ ranges from 0 to 15. $\eta=16$ was used to adjust the width of each Gaussian so that it matches with the separation between two consecutive reference distances. Finally, $f_C(R_{ij})$ is a cutoff function that smoothly modulates the Gaussian term around the cutoff radius, with the following formula and cutoff radius $R_C=5.2$\r{A}:
\begin{equation}
f_C(R_{ij})=
\begin{cases}
(1+cos(\pi\frac{R_{ij}}{R_C}))/2 & R_{ij}\le R_C\\
0 &\text{otherwise} \\
\end{cases}
\label{eq:5}
\end{equation}
We have used 16 distance indices for each atom type and hence 64 radial AEV values. 

Similarly, the angular components of an AEV vector are defined as
\begin{equation}G_{A,B,m,n}^{(ang)}=2^{1-\xi}\sum_{j,k\in\mathcal{N}[i], j\neq k}(1+cos(\theta_{ijk}-\theta_m))^{\xi}f_{(R,n)}(R_{ij},R_{ik})
\label{eq:6}
\end{equation}
\begin{equation}f_{R,n}(R_{ij},R_{ik})=e^{-\eta(R_{ij}+R_{ik})/2-R_n)^2}f_C(R_{ij})f_C(R_{ik})
\label{eq:7}
\end{equation}
with $A$, $B$ defining the two different atom types for nearby atoms, and thus $4+3+2+1=10$ different atom type combinations are possible. $m$ and $n$ are the angle and distance indices that define the reference angles and positions by $\theta_m=\frac{2m+1}{16}\pi$ with $m$ from 0$\sim$7, $R_n=(0.90,1.55,2.20,2.85)$\r{A}, and $\theta_{ijk}$ denotes the angle centered at atom $i$. The same mathematical format of the distance cutoff function was used, but with a radial cutoff value of $R_C=3.5$\r{A}. The normalization constant $\xi=32$. The 10 atom type combinations, 8 reference angles and 4 reference distances altogether defines 320 different angular components of the AEV vector. The calculation of AEVs were performed with the precompiled C++ code from Ref.~\citenum{delta_ML}.

\subsection{NMR shielding calculations and stability analysis} \label{subsec:comp_details}
Recently Liang et al. presented a systematic investigation on using locally dense basis sets (LDBS) and composite QM methods for chemical shieldings calculations, which have been categorized into low-level, middle-level and high-level effectiveness based on a balance of  accuracy and computational cost.\cite{levels} We selected the $\omega$B97X-V functional \cite{mardirossian2014omegab97x} in conjunction with the pcSseg-1 basis set \cite{jensen2015segmented} as our low-level method. The $\omega$B97X-V functional offers robust and transferable performance for various properties prediction,\cite{mardirossian2017thirty,goerigk2017look,hait2018dipole,dohm2018comprehensive,veccham2020density,kim2021establishing,hait2021too,liang2022revisiting} particularly the dipole moment,\cite{hait2018dipole} a  simple but effective measure of electron density in polar molecules. We opted for this functional over the low-level methods recommended in Ref.~\citenum{levels}, which provide more accurate shielding predictions, because those methods could potentially benefit from error cancellation. Thus, we believe it is more advantageous to use $\omega$B97X-V as input for predicting high-level results. The advantage of using $\omega$B97X-V for the low-level input was also validated by its better in-distribution and out-of-distribution predition error when comparing models trained with different low-level methods as input, which are described in Supplementary Table 1. The ORCA 5.0.3 software \cite{neese2020orca} was utilized for these calculations, and local exchange-correlation integrals were computed over DefGrid3, a default ORCA grid, for all atoms. GIAOs\cite{GIAO} were used in all shielding calculations, including subsequent high-level computations.

We directly adopted the high-level method suggested in Ref.~\citenum{levels}, namely CCSD(T)/pcSseg-1 with a basis set correction between pcSseg-1 and pcsSeg-3 calculated from the resolution of identity Møller-Plesset second-order perturbation theory (RIMP2), abbreviated with CCSD(T)(1)$\cup$RIMP2(3). This high-level method can achieve impressively low root mean square errors (RMSEs) (0.048 ppm for H, 0.47 ppm for C, 3.58 ppm for N, and 4.68 ppm for O) in comparison to the theoretical best estimates, CCSD(T) with a complete basis set (CBS). The CFOUR program package, version 2.1, was utilized for CCSD(T) computations \cite{matthews2020coupled,stanton2010cfour,harding2008parallel}, while ORCA was used for RIMP2 calculations. In RIMP2 calculations, the def2-JK \cite{weigend2008hartree} auxiliary basis set was employed for the Coulomb and exchange part, whereas the cw5C \cite{hattig2005optimization} auxiliary basis set was used for auxiliary correlation fitting to expedite the computation.

As our training set encompasses many conformations far from equilibrium and quantum mechanical (QM) calculations are likely to fail, we employed the stability analysis\cite{seeger1977self} at HF/pcSseg-1 level to validate our calculations. We exclude all conformations that might exhibit instabilities, including Restricted HF (RHF) $\rightarrow$ RHF, RHF$\rightarrow$ Unrestricted HF, and RHF $\rightarrow$ Complex RHF.

\subsection{Dataset preparation}
The ANI-1 dataset \cite{ani1}, which contains over 20 million off-equilibrium geometries of small organic molecules up to 8 heavy atom obtained through normal mode sampling, together with the equilibrium structures of these 57,462 molecules, were used to define the most inclusive dataset (DS-ANI-1) used in this work. However, it is very challenging to perform chemical shielding calculations for all the data in DS-ANI-1,  even at a low-level DFT level of theory, and is not accessible for the CCSD(T) calculations that are orders of magnitude more time-consuming than DFT calculations. 

To reduce the size of the dataset while keeping the diversity of the conformations of the molecules, a ``farthest sampling'' algorithm was developed that down-samples off-equilibrium geometries for each molecule in the ANI-1 dataset. The root-mean-square-deviations (RMSDs) for molecules after the optimal alignment using the Quarternions method \cite{quarternion_alignment} was used to evaluate conformation dis-similarities between geometries of the same molecule. A conformation collection pool was defined with the first conformation of a molecule being the first element. In each iteration, the aligned RMSDs for all geometries in ANI-1 dataset but not in the collected pool were calculated towards all conformations in the collected pool, and the geometry with the highest RMSD was added to the collected pool. 

The total number of collected conformations depends on the number of heavy atoms in the molecule. For molecules up to 4 heavy atoms, 200 most dissimilar conformations were collected into the pool. For molecules with 5, 6 and 7 heavy atoms, the number of non-equilibrium conformations collected for each molecule were 100, 50 and 5 respectively. The equilibrium geometries for molecules with 5-7 heavy atoms were always included in the dataset. A stability analysis was performed to further exclude systems for which the NMR shielding calculations are likely to fail or be erroneous. This collection of a sub-sampled dataset (DS-SS) is our primary data for model training and development of the active learning workflow of the iShiftML model, which contains 12,677 geometries for molecules up to 4 heavy atoms, 13,313 geometries for molecules with 5 heavy atoms, 31,462 geometries for molecules with 6 heavy atoms and 37,105 geometries with 7 heavy atoms. 

Using the geometries of all these data, we calculated the DIA and PARA matrix elements under the low-level composite DFT method $\omega$B97X-V/pcSseg-1 DFT.\cite{levels} For the dataset with 5-7 heavy atoms, 1500 geometries were selected from active learning to perform the high-level composite method\cite{levels}. The active learning dataset covering all data using the high-level target values are subsequently labeled DS-AL-N, where N ranges from 4-7, which represents the maximum number of  heavy atoms included in the dataset.  Finally, 41 randomly selected molecules with 8 heavy atoms were collected from DS-ANI-1. The equilibrium geometries and a random non-equilibrium geometry for each of the 41 molecules were used to define our test dataset. Any data point for carbon chemical shielding with significant deviation between calculated low-level and high-level chemical shieldings was excluded. Our full training and testing dataset are provided in the Supplementary Information.

\section{iShiftML Ensemble Model and Training Details}
We have employed an ensemble machine learning approach by randomly splitting the training and validation data into 5 even portions, and 5 separate ML models were trained, each model using a different portion as validation data and the rest as training data. In addition, the network parameters for these five models were also initialized with different random numbers. After all models have been trained, they are combined into an ensemble model. When making predictions, each model in the ensemble predicts a value, and the prediction is given by the average from each of the 5 individual models in the ensemble. 

Because outliers resulting from failed predictions may contaminate the average, any outliers should be identified and excluded from the calculation. To estimate outliers, we used the local outlier factor (LOF) algorithm implemented in the scikit-learn package to detect outliers.\cite{LOF_algorithm} The algorithm relies on a local neighbor density estimation to identify outliers as data points that have a significantly lower density of neigbors than the rest of the data points. Finally, the average and standard deviation among the non-outlier predictions were calculated.

All five ML models are trained by minimizing the mean squared error between the predicted isotropic chemical shieldings and the calculated high-level targets, under the following loss function:
\begin{equation}\mathcal{L}=\frac{1}{N}\sum_{n} (f_{\theta}(X_n)-Y_n)^2
\label{eq:8}
\end{equation}
where $f_\theta$ represents the networks parameterized by $\theta$, $X_n$ are the input features, and $Y_n$ are the target values. Weight decay of $3\times 10^{-5}$ and dropout with probability $0.1$ \cite{dropout} were used after each linear layer to reduce overfitting to the training data. Starting from a learning rate of $1\times 10^{-3}$, a stepwise learning rate decay schedule was used that monitors evaluation performance on the validation dataset, and reduces learning rate by 30\% if the validation error did not decrease after 20 epoch since last error reduction on the validation dataset, unless the learning rate is already smaller than $1\times10^{-6}$. The neural network was implemented in pytorch\cite{pytorch} and optimized using the Adam optimizer\cite{adam} with a batch size of 128 and was trained for 750 epochs.

\section{Results}
\begin{figure}[H]
\begin{center}
\includegraphics[width=0.95\textwidth]{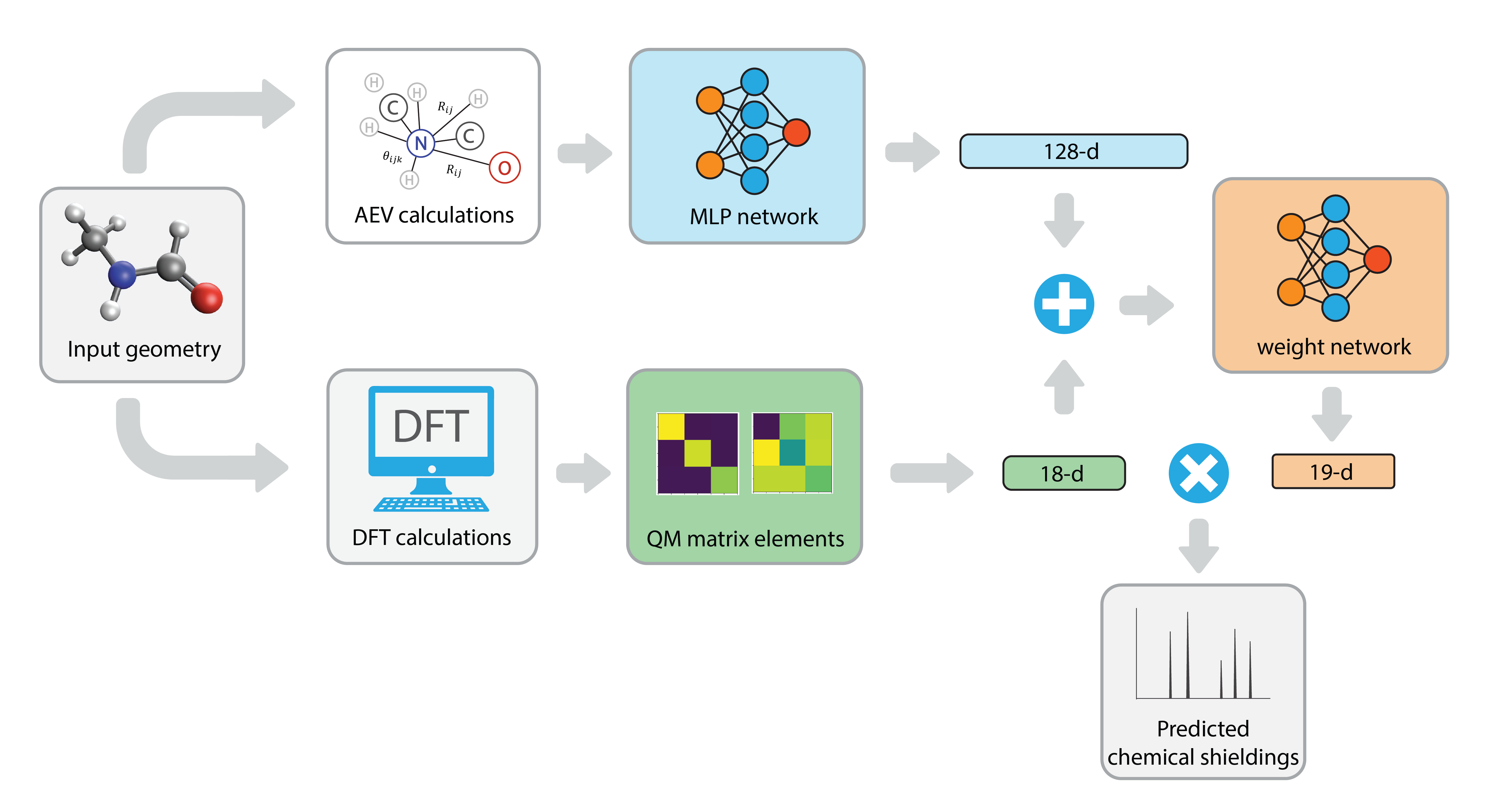}
\end{center}
\vspace{-5mm}
\caption{\textit{The iShiftML ensemble learning model that uses low-level QM calculations of the shielding tensor and AEVs to predict high-level chemical shieldings.} (a) Given a molecular geometry, the AEV around each nucleus is prepared, and are sent into a multi-layer perception (MLP) network with two layers, each of which contains 128 neurons, in which the ReLU activation function\cite{relu} is used for the first layer to encode the AEVs into an 128-dimension internal representation. On a second branch, we perform low-level composite QM calculations to obtain the 18 DIA and PARA chemical shielding values that are concatenated with the AEVs from the first branch to provide input for the second MLP weight network. The weight MLP is composed of a first layer containing 64 neurons and uses ReLU activation, followed by a second layer of 19 neurons and a bias term without an activation function.}
\label{fig:model}
\end{figure}

We begin with the results concerning the iShiftML model itself, the benefits of ensemble training, and a new active learning protocol in order to emphasize the ability to generalize, predict error confidence, and to construct affordable datasets for chemical shift prediction. A schematic of the iShiftML model architecture is depicted in Figure \ref{fig:model}. For a given input geometry, the atomic environment vectors together with the paramagnetic and diamagnetic elements of the shielding tensor are calculated with the lower-level $\omega$B97X-V/pcSseg-1 composite method, and are used as neural network inputs that are trained to predict chemical shieldings of the high-level composite method CCSD(T)(1)$\cup$RIMP2(3) for the four atom types for which we predict chemical shieldings: hydrogen, carbon, nitrogen and oxygen. 

\begin{figure}[H]
\begin{center}
\includegraphics[width=0.9\textwidth]{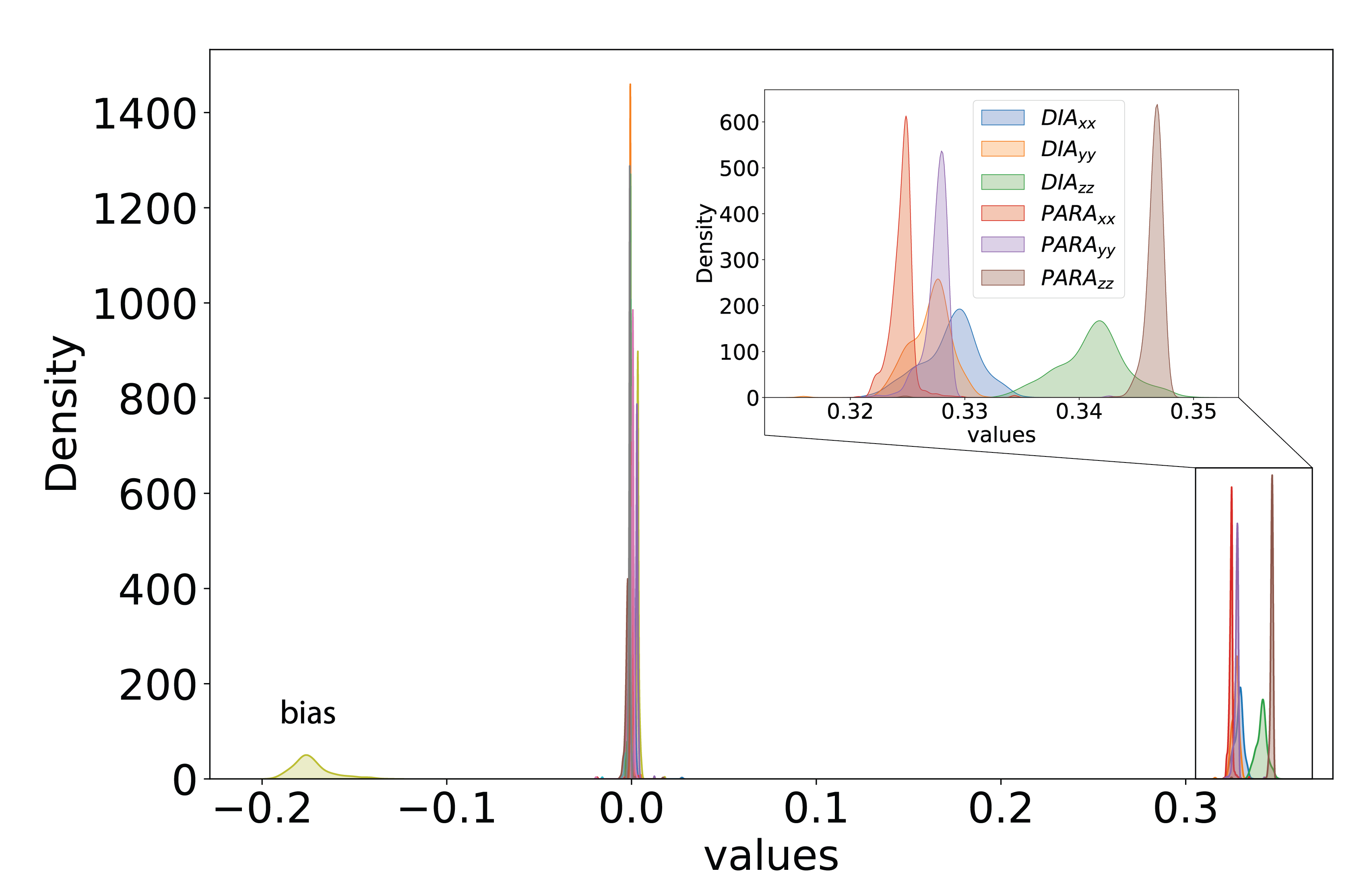}
\end{center}
\vspace{-5mm}
\caption{\textit{Distributions of weight network outputs for hydrogen model evaluated on test data.} Distributions of the weights for diagonal elements in the DIA and PARA matrices are centered close to 1/3, off-diagonal elements are centered around 0, and the bias term is distributed around -0.17.}
\label{fig:weights}
\end{figure}

Figure \ref{fig:weights} shows the distribution of the learnable weights of the 18 DIA and PARA values from the network for the test dataset of the hydrogen atom after the training converges. Even without explicit enforcement, the diagonal elements from the DIA and PARA matrices have weights that are close to $\frac{1}{3}$ and off-diagonal elements distributed around 0, which is consistent with Eq. \ref{eq:3}; the bias term of -0.17 indicates the low-level chemical shieldings have a systematic offset from the more accurate high-level targets. This result proves that the model captures the physical connection between the isotropic chemical shieldings and the intermediate QM matrix elements, and should be generalizable to new predictions even outside of the training dataset, as long as the low-level QM matrix elements are reasonably accurate.

\begin{figure}[H]
\begin{center}
\includegraphics[width=0.95\textwidth]{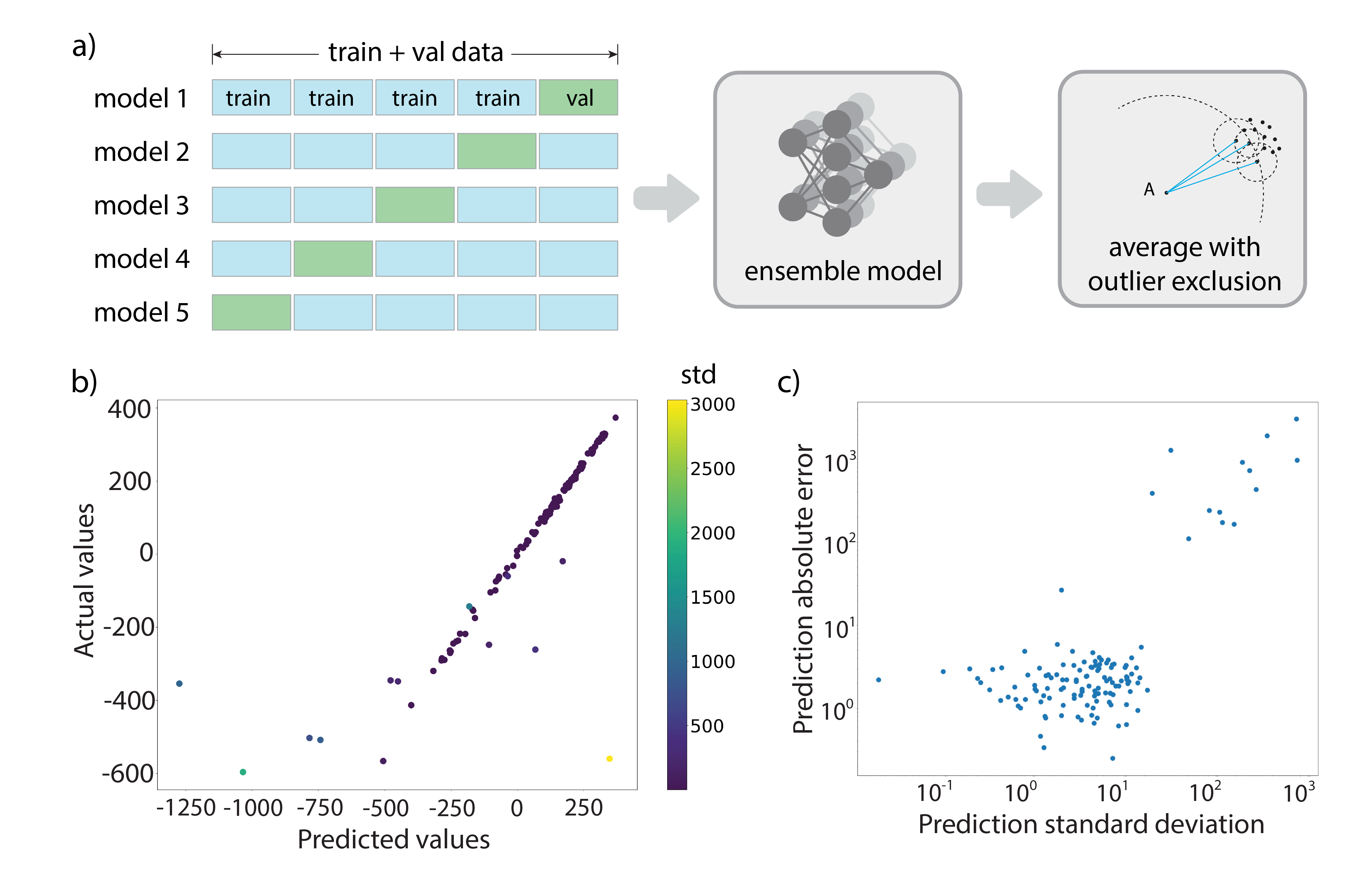}
\end{center}
\vspace{-5mm}
\caption{\textit{Ensemble prediction and correlation with actual prediction error.}(a) An ensemble learning approach using 5-fold cross validation to train individual models in the ensemble. The final prediction is the average prediction from the models after excluding outliers recognized by the Local Outlier Factor algorithm \cite{LOF_algorithm}.(b) An undertrained model for oxygen tested on the 8-heavy-atom test dataset, showing correlation between predicted and actual values. Data points are colored according to their standard deviation (STD), with warm colors representing high STDs and cool colors representing low STDs. (c)  Prediction errors compared to reference values are found to be well correlated with standard deviations of the predictions in the ensemble on a log-log plot. See Methods for further detail.}
\label{fig:ensemble}
\end{figure}

We have also employed an ensemble prediction technique to improve on the accuracy compared to any individual training of the iShiftML model (Figure \ref{fig:ensemble}a). Table \ref{tab:rmse} shows the performance comparisons for individual models and the ensemble average for a model trained with DS-AL-4 for oxygen. We see that while an individual model may make large errors, such as in models 3 and 5, the ensemble average model can mitigate these erroneous predictions, and still reach a consensus prediction that has a lower RMSE and standard deviation than any individual model.

\begin{table}[ht!]
    \caption{Root mean square errors (RMSE) and standard deviations from individual models and from the ensemble model for oxygen prediction when trained using DS-AL-4. Data with high standard deviations (std$>$30) has been excluded to make the trend more concise. All units in ppm. See Methods for further detail.}
    \centering
        \begin{tabular}{l c c }
        \hline
        \hline
        & RMSE & standard deviation \\
        \hline
        Model 1 & 8.30 &5.23 \\
        Model 2 & 8.65&6.18 \\
        Model 3 & 16.76&15.01 \\
        Model 4 & 8.86&5.73 \\
        Model 5 & 23.34&21.72\\
        Ensemble model & 7.60 & 4.82\\
        \hline
        \hline
        \end{tabular}
        \label{tab:rmse}
\end{table}

But just as importantly the ensemble model can provide standard deviations that can be used to estimate actual prediction errors even without knowing the actual ground truth for the chemical shift value. Figure \ref{fig:ensemble}b shows an undertrained model using DS-AL-4 evaluated on 8 heavy atom test data, and compares the predicted and target chemical shielding values with data points colored by the standard deviations from the ensemble. We find that when 

%\subsection{Active learning improves prediction accuracy on test data with larger molecules}
\noindent
the standard deviations are small, the predictions are accurate, and correspondingly all data points with large standard deviations correlate with high predicted errors. Figure \ref{fig:ensemble}c further illustrates the correlation between the prediction standard deviation and the absolute error from the ensemble prediction. Because we find that standard deviations are good approximators for prediction error, the iShiftML model can be applied to and make good prediction for any organic chemical system by only selecting predictions with low standard deviations.

\begin{figure}[H]
\begin{center}
\includegraphics[width=0.95\textwidth]{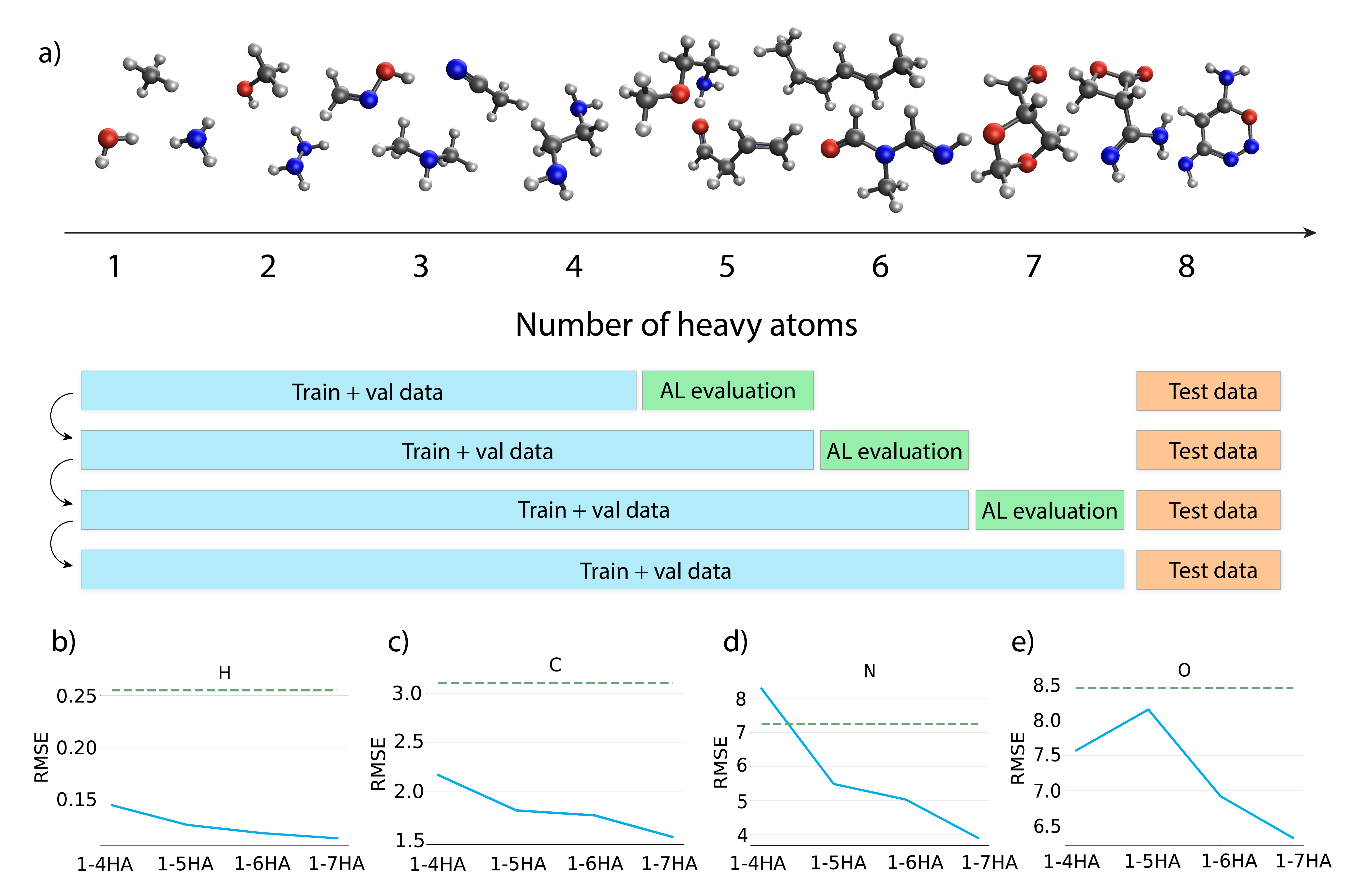}
\end{center}
\caption{\textit{Procedure and results of the active learning workflow.} a) The active learning (AL) workflow. Starting from a model trained with data up to 4 heavy atoms (HA), data with 5HA are evaluated using the trained model and 1500 structures with largest predicted standard deviations from 5HA dataset was included to define the training dataset for the next iteration, until the dataset contains molecules up to 7HA. The 8HA dataset was always used for test. b-e) RMSE on the 8HA test dataset for models trained with AL under training dataset containing molecules with different sizes (blue curve), and also a baseline model that is trained using linear regression (green dotted line). Figures are for hydrogens (b), carbons (c), nitrogens (d) and oxygens (e). (b-e) are also provided in tabular form in Supplementary Table 2. Note that the RMSEs are calculated with uncertain predictions excluded, which removes any prediction with ensemble standard deviation larger than 30. The proportion of data that has been excluded is also listed in Supplementary Table 3.}
\label{fig:active_learning}
\end{figure}

Finally, the ability to identify out-of-distribution data not effectively covered by existing training data through ensemble learning has inspired a novel active learning technique to select only the most important training data to calculate time-consuming high-level chemical shieldings while still improving model performance. In particular given that the high-level QM calculation scales as $O(N^7)$ with system size, it is best to generate as many training data with smaller number of heavy atoms in order to reduce the number of calculations needed for molecules with more heavy atoms (Figure \ref{fig:active_learning}a).

In this case we start by training a model with all subsampled data with up to 4 heavy atoms (DS-AL-4) to allow sufficient initial coverage of the chemical space, and which provides a good starting point for the AL workflow. After training converges with DS-AL-4, the model was used to predict chemical shieldings on the 5 heavy atom data using the low-level QM features. Large standard deviations from the ensemble prediction were utilized to select 1500 structures to generate the next batch of high-level target chemical shieldings which are then added to the training dataset to define the next DS-AL-5 dataset. This process continues until we have included high-level calculations for molecules up to 7 heavy atoms in the training dataset. 

After each AL iteration, the model performance was evaluated on the test dataset composed of randomly selected molecules with 8 heavy atoms to show the effectiveness of the AL approach. Test errors in terms of RMSE for the four atom types are visualized in Figure \ref{fig:active_learning}b-e,  which show the trend of error decrease as larger molecules are added to the training dataset. As a reference, a linear regression (LR) model that uses QM features in DS-AL-7 was also trained, which acts as a baseline equivalent to a model that has fixed coefficients on the DIA and PARA terms instead of atomic environment dependent weights.

Figure \ref{fig:active_learning}b-e shows that even the model trained with DS-AL-4 surpasses the LR reference performance in all atom types other than nitrogen. With more training data included, which we emphasize is that every new training dataset only has $\sim$10\% more data (1500 more molecules) than the dataset with one less heavy atoms, the model continues to systematically improve on the 8 heavy atom test dataset. After the model has been trained with DS-AL-7, the RMSE between predicted and actual high-level QM chemical shieldings are only 0.11 ppm for hydrogen, 1.54 ppm for carbon, 3.90 ppm for nitrogen and 6.33 ppm for oxygen, very close to the error between the target high-level method and the theoretical best estimates. An increase in the proportion of data that can be successfully predicted for this 8 heavy atom test dataset was also observed as more heavy atom data was included in the training dataset (Supplementary Table 3). In the final model trained with DS-AL-7, all test data was predicted with small standard deviations and no data point was excluded from the calculation.

\subsection{Application to predicting gas phase chemical shifts}
The iShiftML model can predict NMR chemical shieldings at a high-level CCSD(T) composite method accuracy using only a tiny fraction of calculation time of a low-level DFT calculation, which enables us to explore new possibilities of experimental CS prediction as well. We first show that experimental gas phase CS for molecules not included in the training dataset can be accurately predicted and error is significantly reduced compared to the low-level DFT that provides the QM matrix elements. Gas phase chemical shifts were used to minimize the effect of environmental complexities, including any influence of solvent and perturbations to chemical shifts due to other molecules nearby. We also consider a more challenging application of the iShiftML method to highlight the transferability of the model for predicting CS for natural products that are much larger and more complex than any molecule in our training dataset. Specifically, calculated CS for 8 diaesteromers of the vannusal B molecules were compared to the experimental measurements to demonstrate that the matching structure can be confidently selected relying on our iShiftML method, which would greatly assist synthetic chemists.

\begin{figure}[H]
\begin{center}
\includegraphics[width=0.98\textwidth]{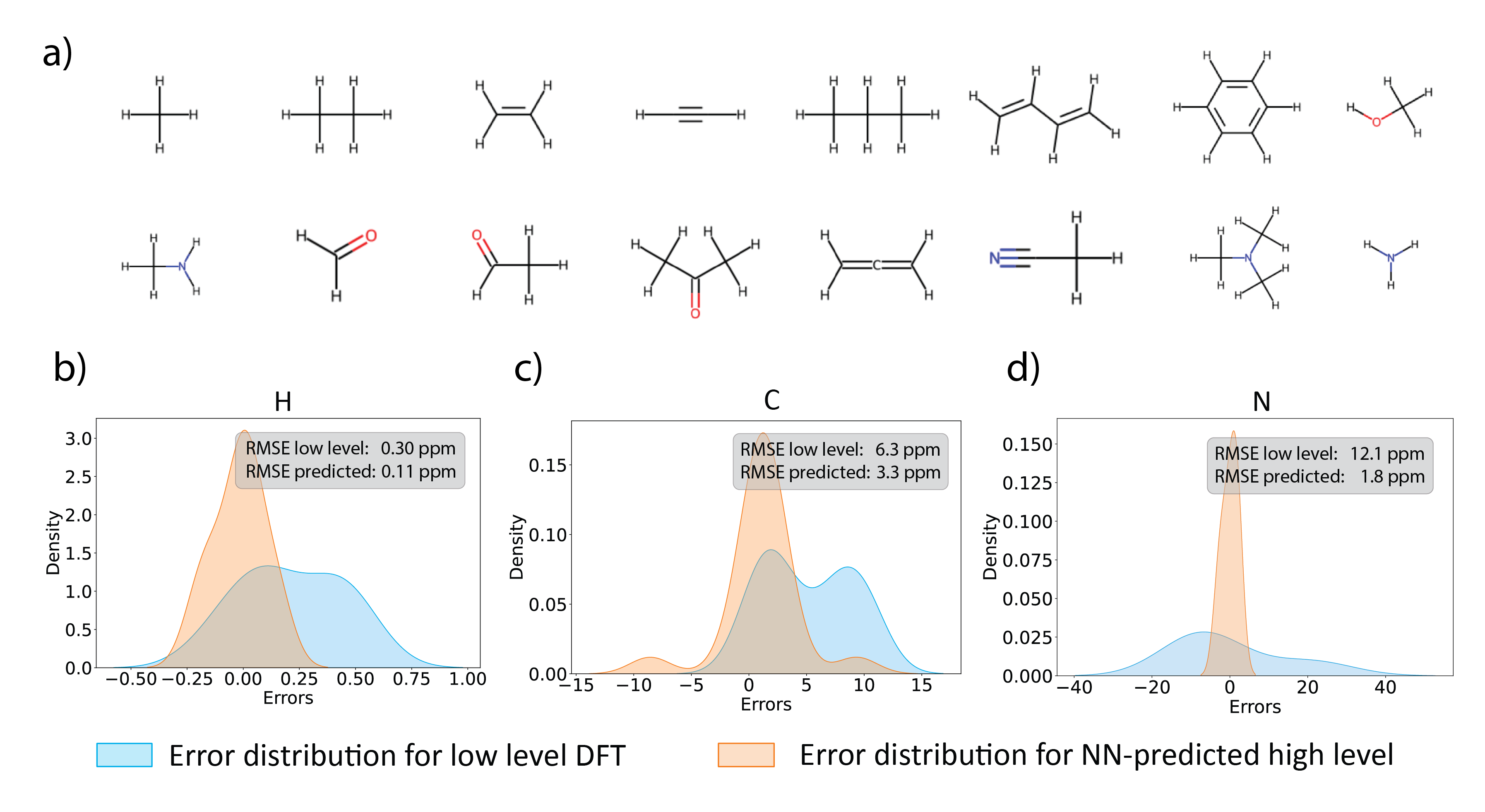}
\end{center}
\caption{\textit{Predicting experimental gas phase chemical shifts for small organic molecules.} (a) the small molecules under investigation. 3D geometries of these molecules are taken from NS372\cite{schattenberg2021extended} and NIST database.\cite{johnson2006nist} (b-d) Distributions of errors between predicted and experimental gas phase NMR chemical shifts for low level DFT calculations ($\omega$B97X-V/pcSseg-1, blue distributions) and iShiftML  predictions for the high level CCSD(T) composite method, orange distributions values for hydrogens (b), carbons (c) and nitrogens (d). Also see Supplementary Figure 1 and Supplementary Table 4.}
\label{fig:gas}
\end{figure}

Figure \ref{fig:gas}a shows a set of 16 molecules that were collected from the literature for their experimental gas phase CS values\cite{exp1,exp2,exp3}, and the geometries of the molecules were taken from NS372\cite{schattenberg2021extended} and NIST database.\cite{johnson2006nist} Because some of these molecules were already in the DS-AL-7 data set, the iShiftML models were retrained after excluding all molecules in Figure \ref{fig:gas}a that are to be tested. 

Chemical shifts were calculated with two techniques. For H and C, the reference chemical shieldings for the respective nuclei in the standard substance tetramethylsilicane (TMS) were calculated at the low-level $\omega$B97X-V/pcSseg-1 and high-level CCSD(T)(1)$\cup$RIMP2(3), and chemical shifts were calculated using $\delta=\sigma_{ref}-\sigma_{nuc}$, where $\sigma_{ref}$ is the isotropic chemical shielding for TMS, and $\sigma_{nuc}$ is the isotropic chemical shielding for the target nucleus. Reference chemical shieldings are 31.766 ppm and 189.588 ppm using the low-level theory for hydrogen and carbon, respectively, while the references were 31.522 ppm and 193.972 ppm using the high-level theory for hydrogen and carbon, respectively. Due to lack of standard substance for nitrogen, a linear model was fit between the predicted chemical shieldings and experimental chemical shifts using a fixed slope of -1 such that only the intercept was fitted. The resulting intercept is -137.9 ppm and -128.3 ppm for the low level and high level theory, respectively. Oxygen nuclei were not assessed due to lack of experimental gas phase chemical shifts for this test set.
 
When compared directly to experimental measurements, we find that iShiftML can predict CS for hydrogen nuclei with RMSE of 0.11 ppm, 3.3 ppm for carbon and 1.80 ppm for nitrogen. By comparison, the low-level DFT calculations gives an RMSE of 0.30 ppm for hydrogen, 6.3 ppm for carbon and 12.1 ppm for nitrogen indicating that with an inexpensive method we have significantly reduced error by 2-6 fold. Figures \ref{fig:gas}b-d show the error distributions for the low-level calculated chemical shifts and high-level predicted chemical shifts, both compared with experimental CS for different nuclei. We see that the low-level CS has a systematic offset for the nuclei under investigation, resulting in error distributions shifted towards positive values for hydrogen and carbon, and negative values for nitrogen. This systematic trend was corrected in the predicted high-level CS, whose errors are centered around zero with a much sharper distribution, in line with its overall superior performance compared to the low-level DFT calculations. One carbon CS (acetylene) that had a high prediction error was also found to have high standard deviation around 13 ppm, which again shows that standard deviations give good estimates of prediction error. 

\subsection{Application to natural product chemical shifts prediction}
\begin{figure}[H]
\begin{center}
\includegraphics[width=0.95\textwidth]{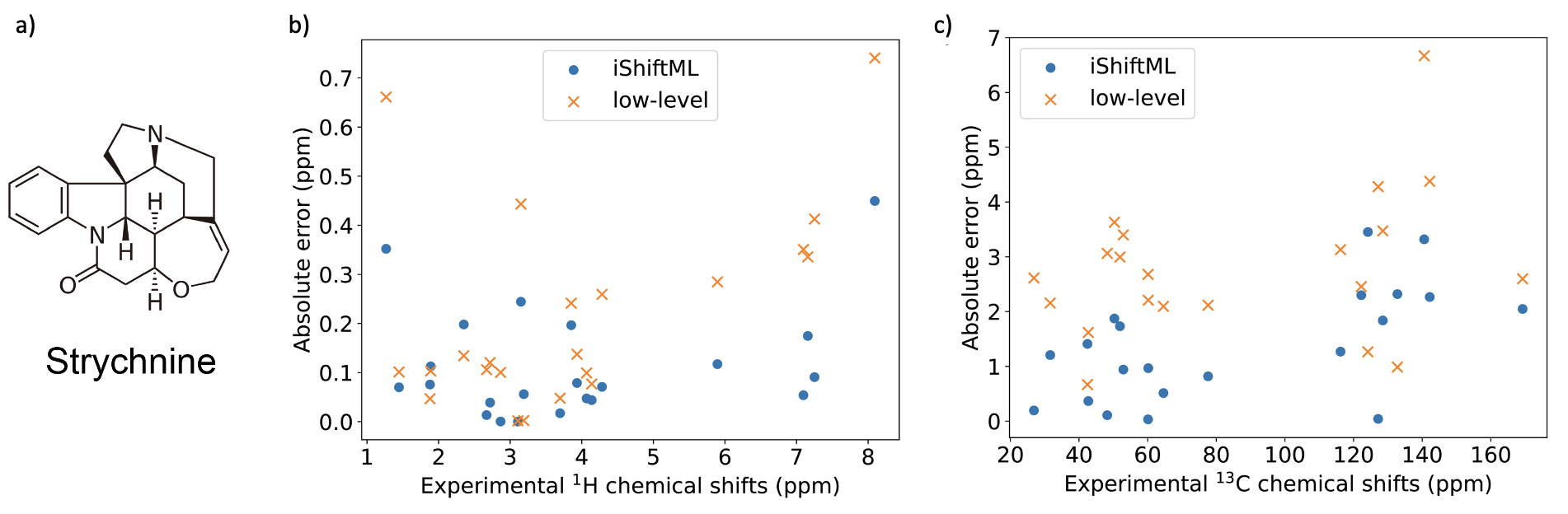}
\end{center}
\caption{\textit{Results on predicting and comparing CS for the strychnine natural product.} a) Molecular structure of strychine. b) Absolute prediction error for the low-level DFT method and iShiftML across the experimental CS range for hydrogens. c) Absolute prediction error for the low-level DFT method and iShiftML across the experimental CS range for carbons. To avoid any inaccuracy in the reference values, all calculated CS were re-referenced to have identical mean values to match the same reference used in the experimental CS. }
\label{fig:natural_product}
\end{figure}

Finally we consider a more challenging application of iShiftML to highlight the transferability of the model. Synthetic chemists often rely on NMR CS as an essential tool to validate the structural correctness of synthesized molecules, especially for natural products.\cite{lodewyk2012computational} In turn, automated methods such as DP4\cite{smith2010assigning} and DP4+\cite{marcarino2022critical} and corresponding ML advances such as DP4-AI\cite{Howarth2020} for computing NMR spectra reliably enough to confirm the chemical composition and stereochemistry of natural products are a critically important counterpart to the experimental data.\cite{bagno2015addressing,semenov2020dft,marcarino2022critical,MacGregor2016} Here we demonstrate that iShiftML can also improve the accuracy of predicted CS for a given molecular structure when compared with experimental measurements. We have used strychnine\cite{strychnine_H,strychnine_C,dft_strychnine,seeman2020decades,semenov2020dft} as a starting example since it is a relatively rigid molecule (Figure \ref{fig:natural_product}a) so that conformational averaging will not play a major role in predicting its chemical shifts accurately. Figure \ref{fig:natural_product}b and c and shows the absolute errors between experimental and calculated CSs using both the low level DFT and high level predictions from the iShiftML model for hydrogens and carbons, and the correlation plots are provided in Figure S2. All iShiftML predictions were made with small standard deviations and hence no outliers were found. 

The RMSEs between experiment and calculated CS for low level DFT, high level iShiftML predictions, along with four other DFT methods reported in Ref. \citenum{dft_strychnine} after re-referencing are also provided in Table \ref{tab:strychnine}. We find that iShiftML has significantly improved over the low level $\omega$B97X-V/pcsSeg-1 DFT calculation that provides input for our model, and is as good or better than other DFT methods that use a much larger basis set. Hence even though strychnine is significantly larger and its fused ring system is not covered by our training dataset, we still realize significant  improvements over much more expensive methods, with errors that remain commensurate with the errors of the 8HA test dataset for high level CCSD(T) calculations. This demonstrates the reliability and generalizability of our model.

\begin{table}[H]

    \caption{
    RMSEs between predicted and measured CS in strychnine using different methods. The 3-dimensional geometry of the strychnine molecule and the experimental measurements of CS are taken from Ref. \citenum{dft_strychnine}. Predicted CS are re-referenced to have same mean values as experimental measurements to avoid any referencing errors. However, the slopes are fixed at unity.
}
\begin{threeparttable}
    \centering
        \begin{tabular}{lll}
        \hline
        \textbf{Method}&\textbf{H}\tnote{b}&\textbf{C}\tnote{c} \\
        \hline
        B3LYP/cc-pVTZ\tnote{a}&\textbf{0.162}&2.095\\
        PBE1PBE/cc-pVTZ\tnote{a}&0.202&2.032\\
        BP/TZP\tnote{a}&0.177&3.145\\
        BP/TZ2P\tnote{a}&0.177&2.895\\
        $\omega$B97X-V/pcsSeg-1 (low-level)&0.296&3.068\\
        iShiftML & \textbf{0.160}&\textbf{1.701}\\
        \hline
        \end{tabular}
        \begin{tablenotes}
        \item[a]{Refitted with unity slope using original data from Ref. \citenum{dft_strychnine}}
        \item[b]{Experimental CS data from Ref. \citenum{strychnine_H}}
         \item[c]{Experimental CS data from Ref. \citenum{strychnine_C}}
        \end{tablenotes}
        
\end{threeparttable}
\label{tab:strychnine}
\end{table}

\begin{figure}[H]
\begin{center}
\includegraphics[width=0.95\textwidth]{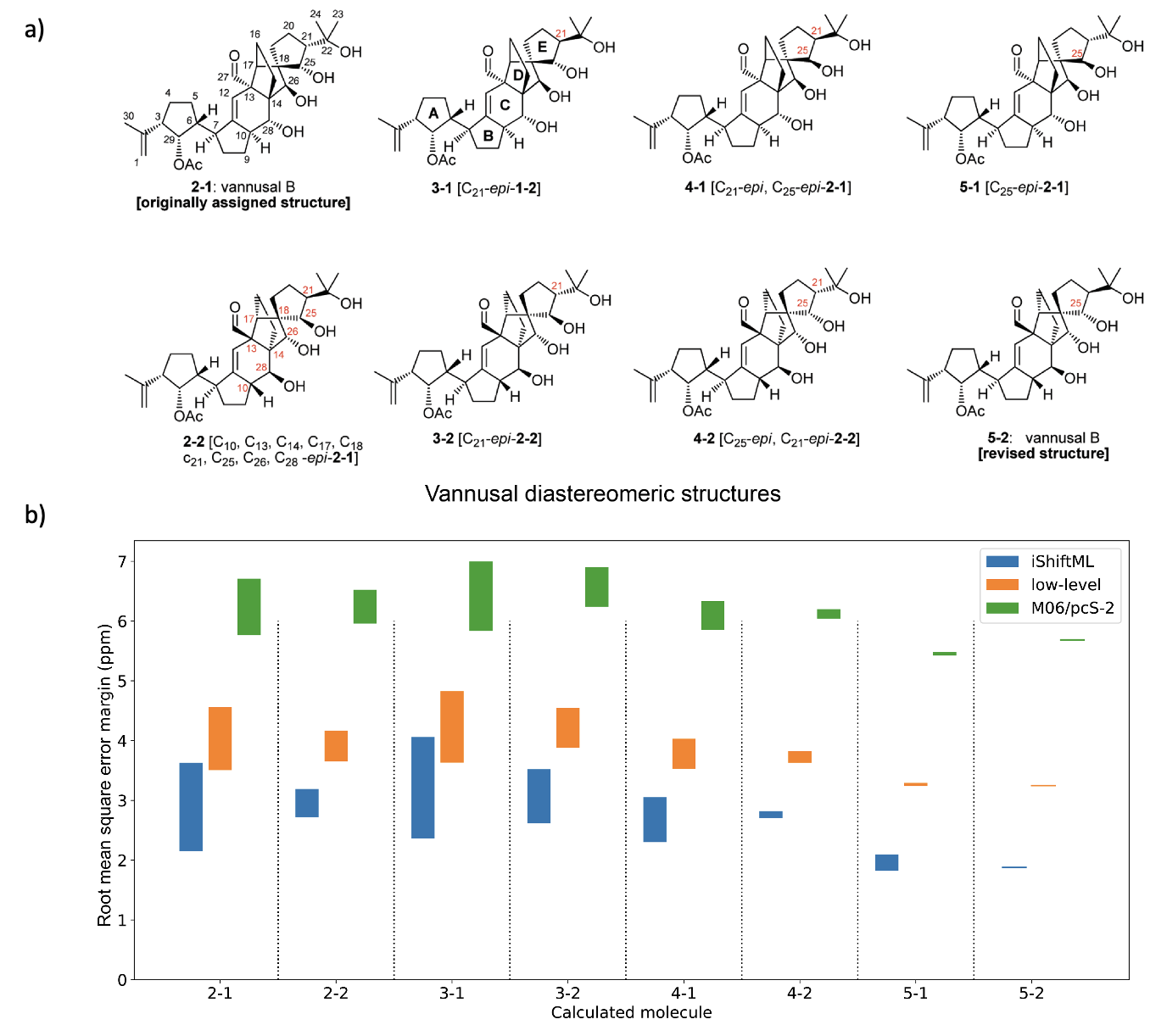}
\end{center}
\caption{\textit{Results on predicting and comparing CS for 8 diastereomers of vannusal B.} a) Molecular structures of the 8 diastereomers of vannusal B. b) Prediction RMSE margins when comparing various vannusal B isomers with their corresponding true experimental CS (bottom position in each bar) and comparing to the vannusal B CS in its native form, 5-2 (top position of each bar) using iShiftML, low-level DFT and M06/pcS-2 (the latter from Ref. \citenum{dft_vannusal}). Large bars with a low bottom  therefore indicate good discrimination between predicted CS for the true structure versus false identification with CS of the native structure. To avoid any inaccuracy in the reference values, all calculated CS were re-referenced to have identical mean values to match the same reference used in the experimental CS. }
\label{fig:natural_product2}
\end{figure}

Finally we consider a more challenging natural product synthesis application to identify the correct molecular structure of  vannusal B (5-2), whose structural assignment had  been uncertain due to the errors in back-calculations and comparison to experiment of a set of highly similar diastereomers of the natural product itself (Figure \ref{fig:natural_product2}a).\cite{vannusal1,vannusal2} Here we have use iShiftML to investigate the match between experimental and calculated CS for carbon atoms, and compare our results with the M06/pcS-2 DFT method reported in Ref. \citenum{dft_vannusal}. However, we did not rescale predicted CS values as was done in Ref. \citenum{dft_vannusal}, so that our reported errors reflect true prediction errors on various atoms in the molecule. Additionally, $sp^2$ hybridized carbons (C1, C2, C11, C12, C21, C31) were retained in our analysis, unlike the original study,  as the iShiftML model should provide accurate predictions (or indicate if it is an outlier) without any prior system knowledge. 

Figure \ref{fig:natural_product2}b provides the RMSEs between predicted and experimental CSs for vannusal B (5-2) and same for the structures of the other diastereomers (2-1, 2-2, 3-1, 3-2, 4-1, 4-2, and 

\noindent
5-1). We find that iShiftML consistently predicts lower RMSE across all molecules compared with the low-level DFT method or M06/pcS-2 from Ref. \citenum{dft_vannusal} (i.e. the bottom of the blue bars for iShiftML are well below the bottom of the orange and green bars). Furthermore, in Figure \ref{fig:natural_product2}b the bottom of each bar provides the RMSE between the experimental chemical shifts that match the true structure of each diastereomer, while the top position of the RMSE bar shows the error made if the experimental CS for natural product structure 5-2 involved an (erroneous) assignment to the diastereomer structure of interest. On average, iShiftML has a larger RMSE margin (longer bars) between the correct structure assignment of the given diastereomer and the erroneous matching (to 5-2) based on the two sets of experimental CS. Therefore iShiftML can identify the correct structure from other candidates with higher confidence, as well as recognize the true vannusal B molecule with ease.

\section{Conclusion}
Methods for \textit{ab initio} calculation of chemical shieldings lie on a spectrum, with one end being DFT calculations that are cheap but less accurate, and the other end being CCSD(T)/CBS methods that are highly accurate but prohibitively expensive for large systems. We have now created a tool to bridge the two ends using machine learning, so that with input features coming from a relatively fast DFT calculation, the predictions can approach the highest level of accuracy achieveable through quantum mechanics calculations, without incurring extra cost. By utilizing a feature set that relies on chemical shielding DIA and PARA tensor components, together with features that describe molecular geometry, we demonstrated that iShiftML can achieve not only excellent accuracy compared to the high level target chemical shieldings, but greater transferability to test molecules larger than any molecule contained in our training dataset, approaching the intrinsic errors for the high level targets when compared to CCSD(T)/CBS calculations. 

While iShiftML is readily helpful for those who study the chemical shieldings of small organic molecules using coupled cluster methods, its broader applicability is exemplified with predicting experimental chemical shifts with higher accuracy for arbitrary systems. Our trained model without any fine-tuning can predict gas phase experimental chemical shifts for small organic molecules with excellent accuracy and reduce error by more than 50\% compared to the direct calculation using the same level of QM theory as our input features. When applying this method to synthesized natural products, we illustrated it could achieve better agreement between predicted and measured chemical shifts when the structures match, and provide better differentiation capability between matched and mismatched diastereomer structures given the CS experimental data. We believe there are many more application possibilities of our method, including predicting chemical shifts for proteins, correcting assignment errors in databases, and aiding drug discovery in determining structure-activity relationships.

There are also some limitations of the current method. It is trained with equilibrium and non-equilibrium geometries of closed-shell small organic molecules that contain only H, C, N and O atoms. Also, only single molecule data were included in our training dataset. Therefore it is not expected to work for open-shell molecules, molecules containing other elements, or for molecular systems in which intermolecular interactions play a major role in the chemical shifts. However, we are planning to improve the method in the future to make it even more transferable and widely applicable. For example, adding support for more atom types will be our first step to allow this method to work for a broader range of organic compounds. Nevertheless, we believe in its current form iShiftML can already benefit those in need of a fast and reliable chemical shift predictor. 

% For example, we envision this application as a pilot study in predicting chemical shifts for PTMs for proteins but more needs to be considered. Specifically, the analysis only used individual structures from NMR ensembles, thereby neglecting the dynamic nature of proteins in solution. The tripeptide-fragments of residues are an incomplete picture of the local environments of nuclei under investigation, which ignores long-range and non-bonded interactions. Furthermore, only PTM structures of the discussed proteins were available, therefore the biopolymers before PTM have to be modelled manually, and any potential conformational changes incurred by PTMs are neglected. Future work is underway to develop this method to a complete CS back-calculator for arbitrary PTMs in proteins. 

\begin{acknowledgement}
This work was supported by funding from the National Institute of General Medical Sciences under grant number 5U01GM121667. This research used computational resources of the National Energy Research Scientific Computing Center, a DOE Office of Science User Facility supported by the Office of Science of the U.S. Department of Energy under Contract No. DE-AC02-05CH11231. A. L. Ptaszek was funded by the Christian Doppler Laboratory for High-Content Structural Biology and Biotechnology, Austria and received further support from the DosChem doctoral school program faculty of Chemistry, University of Vienna.
\end{acknowledgement}

\begin{suppinfo}
Scatter plots and tabulated values of experimental chemical shifts versus the predicted or calculated chemical shieldings under the low-level DFT calculation and high-level neural network prediction with ensemble standard deviation. The code package is provided through GitHub at https://github.com/THGLab/iShiftML/.
\end{suppinfo}

\bibliography{references}

\end{document}


\section{Dataset and code links}
GitHub repository link: \href{https://github.com/THGLab/iShiftML}{https://github.com/THGLab/iShiftML}

\noindent DS-SS (subsampled dataset from ANI-1 with unstable molecules excluded): 

\noindent \href{https://github.com/THGLab/iShiftML/blob/master/dataset/DS-SS.txt}{https://github.com/THGLab/iShiftML/blob/master/dataset/DS-SS.txt}

 \noindent DS-AL (active learning dataset): 
 
 \noindent\href{https://github.com/THGLab/iShiftML/blob/master/dataset/DS-AL.txt}{https://github.com/THGLab/iShiftML/blob/master/dataset/DS-AL.txt}

Removed chemical shielding: 8\_atom/mol\_34274/99.xyz/atom\_6 (calculated low level chemical shielding:  -2.066, calculated high level chemical shielding: 197.792)

\section{Supporting Figures}
\begin{figure}[H]
\begin{center}
\includegraphics[width=0.98\textwidth]{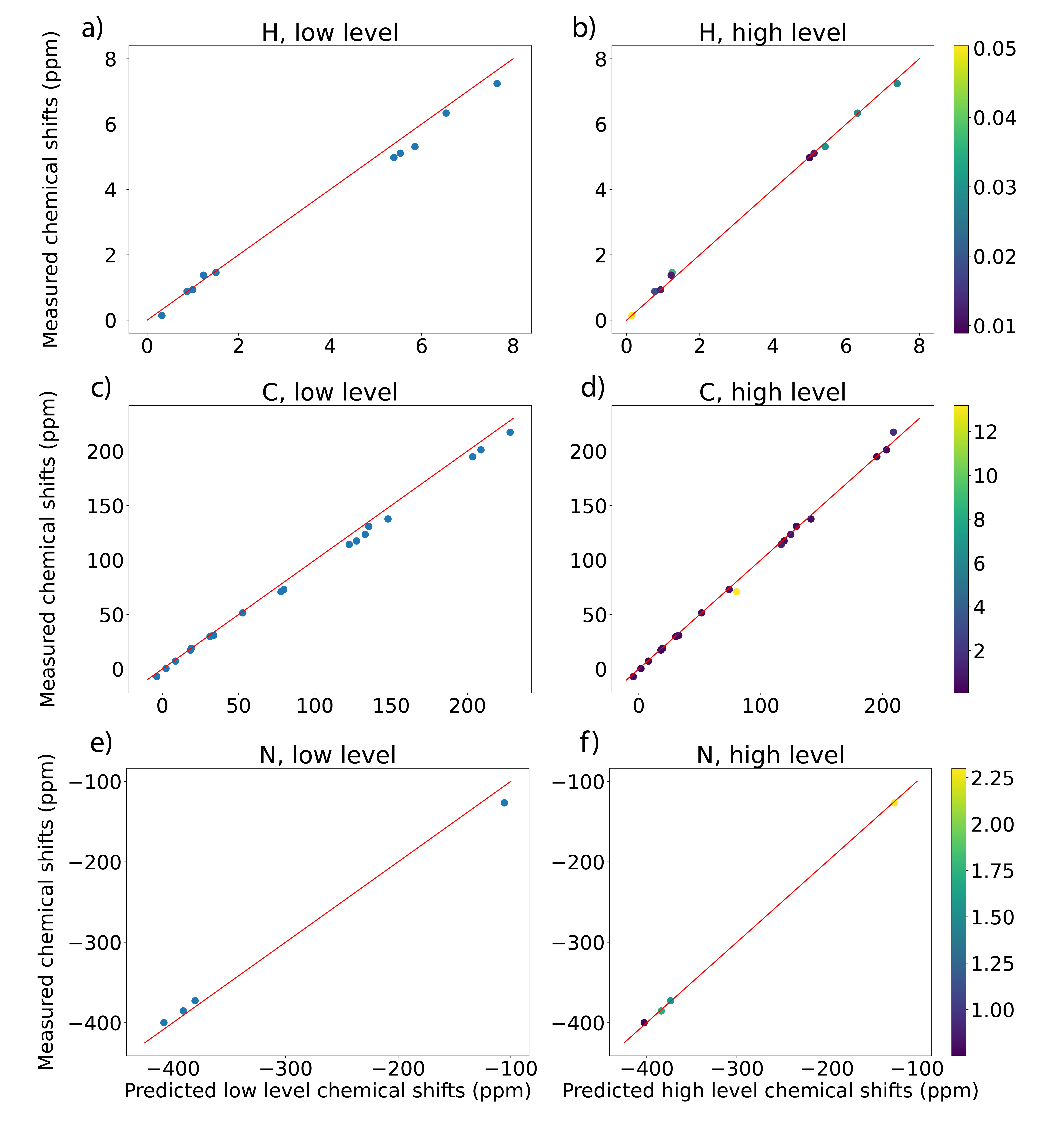}
\end{center}
\caption{\textit{Scatter plots on low-level and iShiftML predicted high-level chemical shifts compared to experimentally measured gas phase chemical shifts for different atom types.} a), c) and e) show the calculated chemical shifts using low level $\omega$B97X-V/pcSseg-1 DFT methods, while b), d) and f) show the predicted chemical shifts using iShiftML targeting CCSD(T)(1)$\cup$RIMP2(3) composite method accuracy, with data points colored by the standard deviation from the ensemble using color codes on the right. Red lines in the figures represent y=x.}
\label{fig:s1}
\end{figure}

\begin{figure}[H]
\begin{center}
\includegraphics[width=0.98\textwidth]{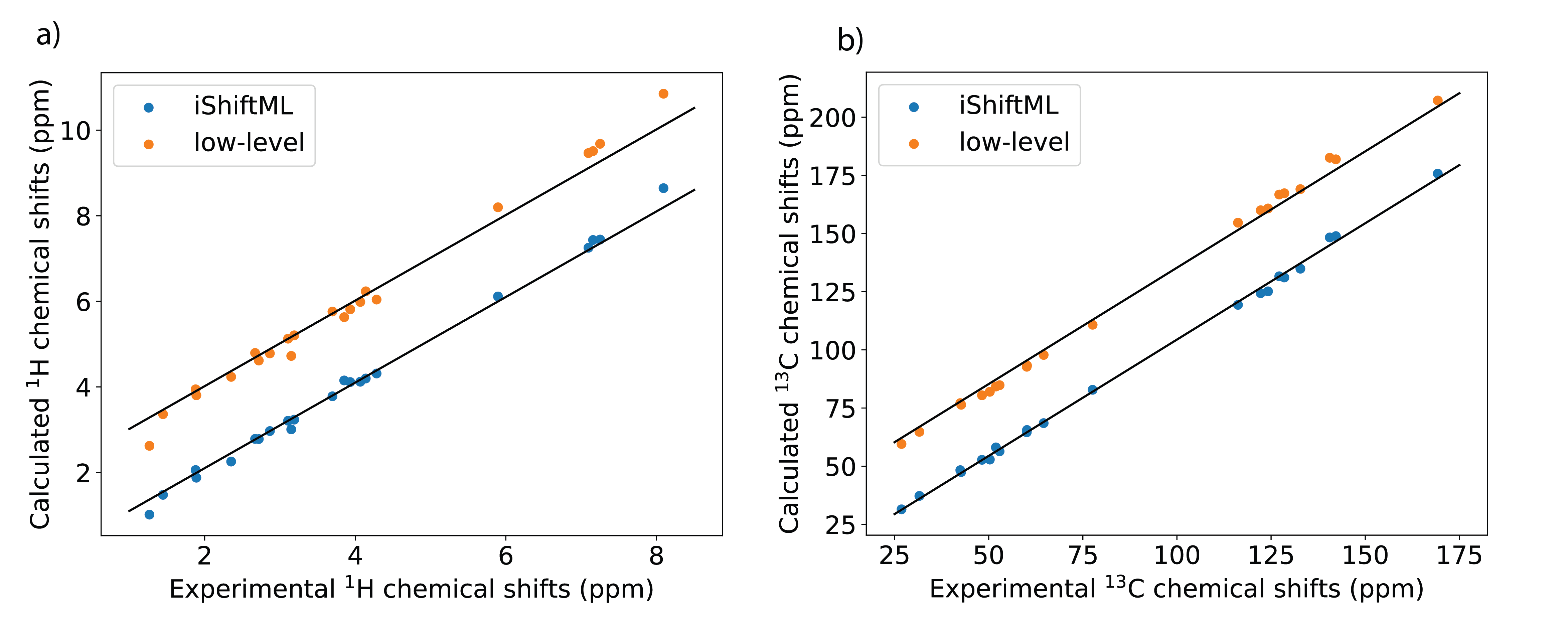}
\end{center}
\caption{\textit{Scatter plots on calculated chemical shifts compared to experimental chemical shifts using low-level DFT and iShiftML for strychnine.} a) Scatter plots for hydrogens. b) Scatter plots for carbons. Black lines represent y=x. Data for low-level DFT are shifted on the y axis for 2 ppm (a) and 30 ppm (b) to make comparisons more clear.}
\label{fig:s2}
\end{figure}

%\begin{figure}[H]
%\begin{center}
%\includegraphics[width=0.98\textwidth]{figures/figure_s_corr.png}
%\end{center}
%\caption{\textit{RMSE heatmaps for pairwise comparisons between experimental and calculated chemical shifts for the 8 diastereomers of vannusal B.} Each cell is colored by the RMSE between chemical shifts predicted by different methods with the molecule indicated on the x axis and the experimental chemical shifts with the molecule indicated on the y axis. Calculated chemical shifts are re-referenced to have identical mean values as experiment and no chemical shifts are excluded in the calculation of RMSE values. a)Our method (iShiftML), b) Low-level DFT method ($\omega$B97X-V/pcsSeg-1), c) Method from Ref.\citenum{dft_vannusal} (M06/pcS-2).}
%\label{fig:s3}
%\end{figure}

\begin{figure}[H]
\begin{center}
\includegraphics[width=0.98\textwidth]{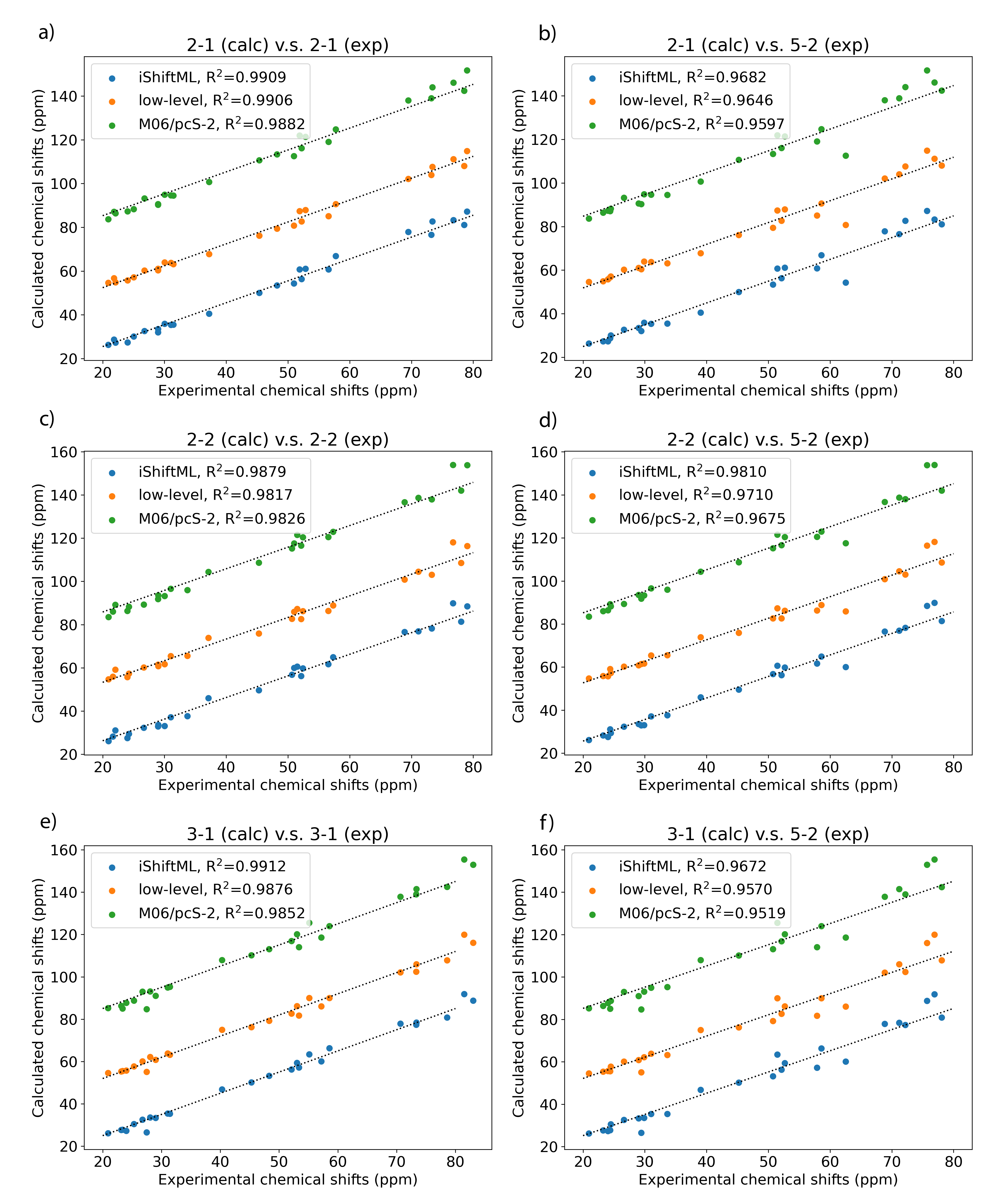}
\end{center}

\end{figure}

\begin{figure}[H]
Figure 3 (continued)
\begin{center}
\includegraphics[width=0.98\textwidth]{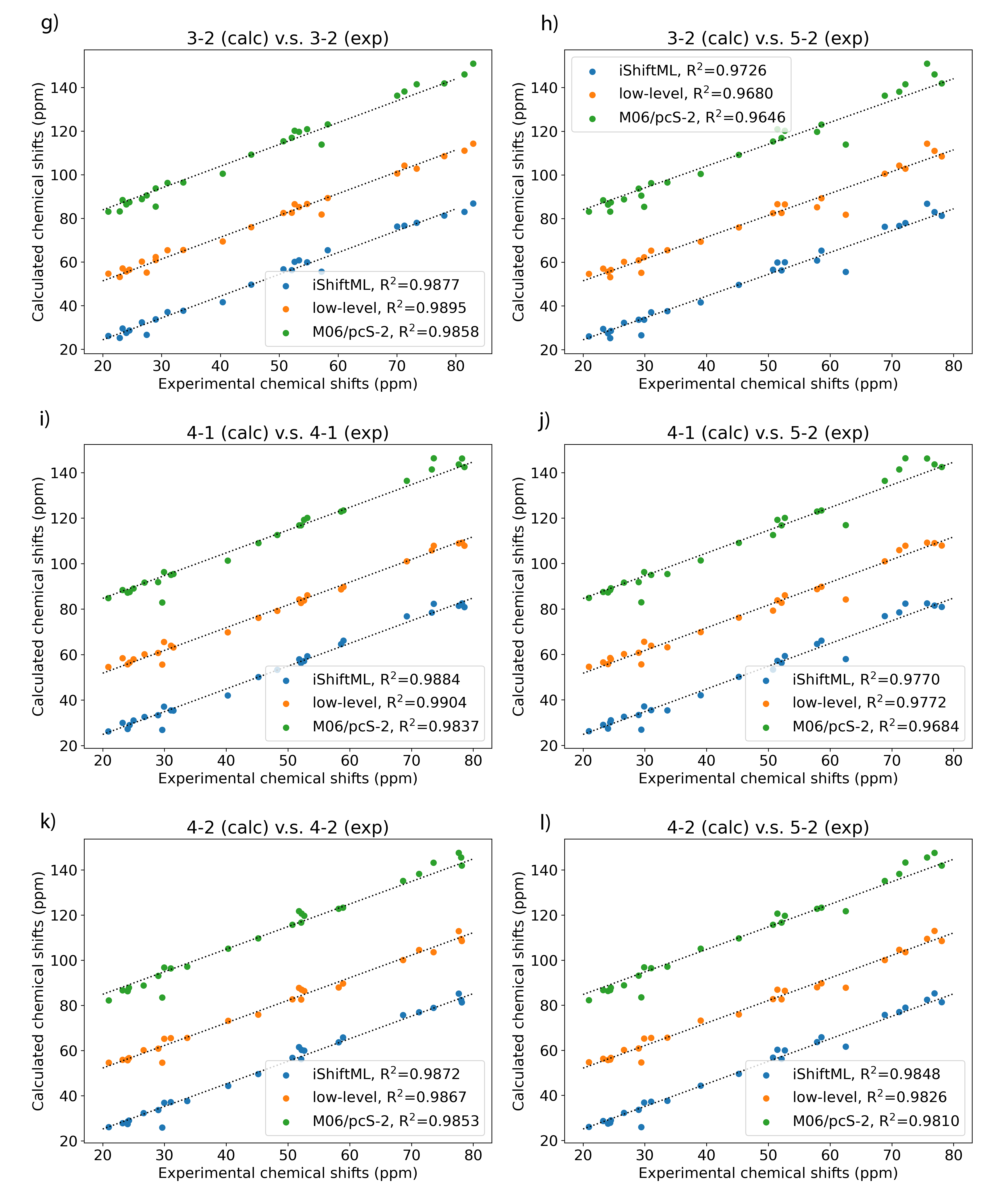}
\end{center}

\end{figure}

\begin{figure}[H]
Figure 3 (continued)
\begin{center}
\includegraphics[width=0.98\textwidth]{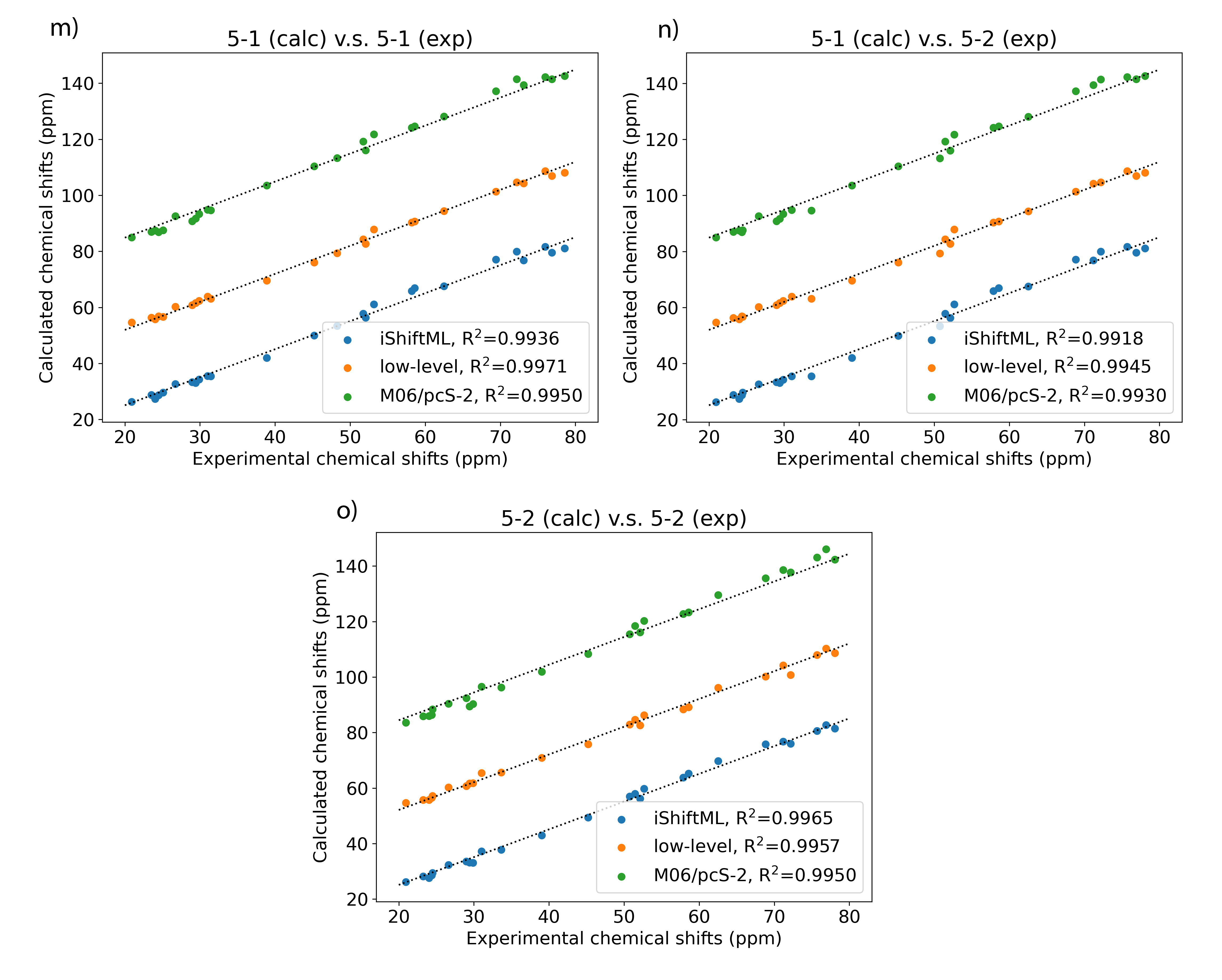}
\end{center}
\caption{Scatter plots on calculated carbon chemical shifts compared to experimental chemical shifts using low-level DFT, iShiftML and M06/pcS-2 for pairs of comparison between diastereomers of vannusal B. }
\label{fig:s4}
\end{figure}
\clearpage

\section{Supporting Tables}

\begin{table}[h]
    \centering
    \caption{Prediction root mean square errors (RMSE) on in-distribution dataset (ID, randomly selected non-equilibrium geometries from training dataset containing up to 5 heavy atoms which are excluded from training), and out-of-distribution dataset (OOD, 5 heavy atom dataset with equilibrium geometries) from models trained using QM features from different low-level methods using pcSseg-1 basis sets. All units in ppm.}
    \begin{tabular}{lcccccccc}
        \hline\hline

\multirow{2}{*}{\textbf {Low level QM method }} & \multicolumn{2}{c}{\textbf { H }}  & \multicolumn{2}{c}{\textbf { C }} & \multicolumn{2}{c}{\textbf { N }} & \multicolumn{2}{c}{\textbf { O }} \\
& \textbf{ID}&\textbf{OOD}& \textbf{ID}&\textbf{OOD}& \textbf{ID}&\textbf{OOD}& \textbf{ID}&\textbf{OOD}\\
\hline
B97-D\cite{grimme2006semiempirical} &0.05 & 0.14 & 0.78 & 2.97 & 2.4&6.0&6.8&12.7 \\
B97M-V\cite{mardirossian2015mapping}  & 0.05 & 0.14 & 1.08 & 2.88 & 2.3&6.3&6.0&13.0 \\
KT3\cite{keal2004semiempirical}  & 0.05&0.14&0.88&2.96&2.5&7.2&7.6&13.0 \\
SCAN\cite{sun2015strongly} & \textbf{0.04}&0.14&0.66&2.81&1.4&5.1&4.3&56.6 \\
$\omega$B97X-V\cite{mardirossian2014omegab97x} &  \textbf{0.04}& \textbf{0.12}& \textbf{0.34}& \textbf{2.66}& \textbf{1.0}& \textbf{4.0}& \textbf{4.2}& \textbf{10.1} \\

        \hline
    \end{tabular}

\label{tab:active_learning_rmse}
\end{table}
\begin{table}[h]
    \centering
    \caption{RMSE on the 8 heavy atom test dataset for models trained with active learning under training dataset containing molecules with different sizes, and also a baseline model that is trained using linear regression (LR). Models trained with DS-AL-N are called 1-N HA models in the table. Outliers (data with predicted standard deviations$>$30) are excluded from the RMSE calculation, and the proportion of outliers is provided in Supplementary Table 3}
    \begin{tabular}{lccccc}
        \hline\hline

\textbf {Model type } & \textbf{Training dataset size} & \textbf { H } & \textbf { C } & \textbf { N } & \textbf{O} \\ \hline
LR(baseline) & 17178 & 0.255 & 3.11 & 7.26 & 8.46 \\
1-4HA  & 12677 & 0.144 & 2.17 & 8.30 & 7.57 \\
1-5HA  & 14177 & 0.125 & 1.81 & 5.49 & 8.15 \\
1-6HA & 15676 & 0.117 & 1.76 & 5.03 & 6.92 \\
1-7HA & 17178 & 0.112 & 1.54 & 3.90 & 6.33 \\

        \hline
    \end{tabular}

\label{tab:active_learning_rmse}
\end{table}

\begin{table}[h]
    \centering
   \caption{Proportion of outliers (data with predicted standard deviations$>$30) for chemical shielding predictions on the 8 heavy atom test dataset using different models during AL training. }
    \begin{tabular}{lcccc}
        \hline\hline

\textbf {Model type } & \textbf { H } & \textbf { C } & \textbf { N } & \textbf{O} \\ \hline
1-4HA  & 0 & 0 & 2.1\% & 10.1\%\\
1-5HA  & 0 & 0 & 1.4\% & 3.9\%  \\
1-6HA & 0 & 0 & 0.7\% & 0.7\%  \\
1-7HA & 0 & 0 & 0 & 0 \\
                                       
        \hline
    \end{tabular}

\label{tab:meanstd}
\end{table}

\begin{table}[h]
    \centering
   \caption{Calculated and predicted NMR chemical shifts for small organic molecules and comparison to experimental gas phase chemical shift measurements. All units in ppm.}
    \centerline{
    \begin{tabular}{lcccc c}
        \hline\hline

\textbf {Molecule } & \textbf { Atom } & \textbf { \shortstack{Experiment \\chemical shifts} } & \textbf { \shortstack{Calculated \\low level \\chemical shifts} } & \textbf{\shortstack{Predicted \\high level\\ chemical shifts}}&\textbf{\shortstack{Predicted \\standard \\deviations}} \\ \hline
CH4  & H & 0.14 & 0.32 & 0.15&0.05\\
C2H6  & H & 0.88 & 0.87 & 0.77 &0.02 \\
C2H4 & H & 5.31 & 5.86 & 5.43 &0.03 \\
C2H2 & H & 1.46 & 1.51 & 1.25&0.04 \\
C3H8 & H (CH3) & 0.93 & 1.00 & 0.93&0.01\\
C3H8 & H (CH2) & 1.38 & 1.23 & 1.22&0.01\\
Butadiene & H (CH2, trans) & 4.98 & 5.40 & 5.00&0.009 \\
Butadiene & H (CH2, cis) & 5.11 & 5.53 & 5.13&0.01 \\
Butadiene & H (CH) & 6.34 & 6.54 & 6.31 &0.03\\
Benzene & H & 7.24 & 7.65 & 7.39&0.03\\
        \hline
        CH4 & C & -7.0 & -3.7 & -4.4 & 0.4\\
        C2H6 & C & 7.2 & 8.7 & 7.9  & 0.2\\
        C2H4 & C & 123.6 & 133.0 & 124.6 & 0.9 \\
        C2H2 & C & 70.9 & 77.8 & 80.2&13.2 \\
        C3H8 & C (CH3) & 17.3 & 18.3 & 18.2 & 0.1 \\
        C3H8 & C(CH2) & 19.0 & 18.9 & 19.6 & 0.1\\
        Butadiene & C (CH2) & 117.5 & 127.3 & 119.2 & 0.3 \\
        Butadiene & C (CH) & 137.7 & 148.0 & 141.2 & 0.4\\
        Benzene & C & 130.9 & 135.3 & 129.3 & 0.5 \\
        CH3OH & C &51.5 &52.7 & 51.6 & 0.5 \\
        CH3NH2 & C & 29.8 & 31.2 & 30.4 & 0.2 \\
        CH3CHO & C (CH3) & 30.9 & 33.7 & 32.8 &0.4\\
        CH3CHO & C (CHO) & 194.8 & 203.5 & 195.2 & 0.2\\
        CH3COCH3 & C (CH3) & 30.1 & 31.6 & 31.2 & 0.3\\
        CH3COCH3& C (CO) & 201.2 & 208.9 & 203.0 & 0.3\\
        CH2CCH2 & C (CH2) & 72.9 & 79.6 & 74.1 & 0.8 \\
        CH2CCH2 & C & 217.4 & 228.1 & 208.8 & 1.9 \\
        CH3CN & C (CH3) & 0.4 & 2.4 & 1.8 &0.2\\
        CH3CN & C (CN)&114.3 & 122.6 & 117.0 & 1.2\\
        \hline
        CH3NH2 & N & -385.4 & -390.9 & -383.9 & 1.7\\
        CH3CN & N & -126.7 & -105.8& -124.9 & 2.3 \\
        N(CH3)3 & N &-372.8 & -380.4 & -373.4 & 1.6\\
        NH3 & N & -400.1 & -408.0 & -402.8 & 0.8\\
        \hline
        \hline

    \end{tabular}
}
\label{tab:meanstd}
\end{table}




\clearpage
\bibliography{references}